\documentclass[jcp,reprint,amsmath,amssymb,floatfix]{revtex4-1}
\usepackage{graphicx}
\usepackage{bm}
\usepackage{amsfonts,mathrsfs} 
\newcommand\tabref[1]{Table~\ref{#1}}

\newcommand{\figref}[1]{Figure~\ref{#1}}
\newcommand{\eqnref}[1]{Eq.~(\ref{#1})}

\begin{document}

\title{Long-range correlation energy calculated from coupled atomic response functions}
\author{Alberto Ambrosetti$^1$}
\author{Anthony M. Reilly$^1$}
\author{Robert A. DiStasio Jr.$^2$}
\author{Alexandre Tkatchenko$^1$}
\email{tkatchen@fhi-berlin.mpg.de}
\affiliation{$^1$Fritz-Haber-Institut der Max-Planck-Gesellschaft, Faradayweg 4-6, 14195, Berlin, Germany \\
$^2$Department of Chemistry, Princeton University, Princeton, NJ 08544, USA}

\date{\today}

\begin{abstract}
An accurate determination of the electron correlation energy is an essential prerequisite for describing the structure, stability, and 
function in a wide variety of systems, ranging from gas-phase molecular assemblies to condensed matter and organic/inorganic interfaces. 
Even small errors in the correlation energy can have a large impact on the theoretical description of chemical and physical
properties in molecular systems of interest. The development of efficient approaches for the accurate calculation of the long-range correlation energy 
(and hence the dispersion energy as well) is essential and such methods can be coupled with many density-functional approximations (DFA), local methods 
for the electron correlation energy, and even interatomic force fields. While a number of methods have been developed to augment DFA via corrections 
for the dispersion energy, most of these approaches ignore the intrinsic many-body nature of correlation effects, leading to inconsistent and sometimes 
even qualitatively incorrect predictions. In this work, we build upon the previously developed many-body dispersion (MBD) framework, which is intimately 
linked to the random-phase approximation (RPA) for the correlation energy. We separate the correlation energy into short-range contributions that are 
modeled by semi-local functionals and long-range contributions that are calculated by mapping the complex all-electron problem onto a set of atomic 
response functions coupled in the dipole approximation. We propose an effective range-separation of the coupling between the atomic response functions that
extends the already broad applicability of the MBD method to non-metallic materials with highly anisotropic responses, such as layered nanostructures.
Application to a variety of high-quality benchmark datasets illustrates the accuracy and applicability of the improved MBD approach, 
which offers the prospect of first-principles modeling of large structurally complex systems with an accurate description of the long-range correlation energy.
\end{abstract}

\maketitle

\section{Introduction}

Density-functional theory (DFT) has emerged as a powerful electronic-structure technique that is increasingly being applied to many areas of chemistry, physics, and materials science. This success can be attributed to the very favorable ratio of computational cost to accuracy in DFT; by offering ``correlation at a mean-field price'' DFT allows for systematic first-principles investigations for relatively large-scale systems.~\cite{Burke_2012} DFT has a much higher computational efficiency compared to wavefunction-based methods, following directly from the mapping between the wavefunction and the electron density as provided by the first Hohenberg-Kohn (HK) theorem.~\cite{HK} In fact, HK proved that the ground-state (gs) energy of a many-electron system can be expressed as a functional of the ground-state electronic density $n(\mathbf{r})$, \textit{i.e.}, $E_\text{gs}=E[n_\text{gs}(\mathbf{r})]$.

Despite the exactness of the HK approach, Kohn-Sham DFT~\cite{KS} relies in practice on an approximate treatment of the exchange and correlation (xc) energy, which
encodes all of the non-trivial quantum-mechanical interactions between electrons. As the xc contribution is essential in order to describe bonding 
and electronic structure, the success or limitation of a given density-functional approximation (DFA) strongly depends on the underlying xc functional.

One of the most widely used classes of approximate xc functionals are the semi-local functionals, which use the density and its gradients to determine the exchange and correlation energy. Due to the potentially quasi-linear scaling of the computational cost (wherein thousands of atoms can now be treated in the most efficient implementations), semi-local DFT represents a work-horse method for molecular and solid-state applications. However, current semi-local functionals have a number of well-known deficiencies, such as self-interaction or delocalization errors~\cite{Cohen_2008} and a lack of long-range electron correlation.~\cite{Klimes_2012} The self-interaction error stems from the approximate treatment of electron exchange. On the other hand, the lack of long-range correlation is a consequence of the semi-local nature of the correlation, which implies an exponential decay of the interaction among separated (charge-neutral) density fragments. This approximation completely neglects long-range correlation due to collective electronic fluctuations, which are known to occur in all molecules and solids. Hence, standard semi-local functionals neglect the long-range dispersion interactions that are essential for cohesion and function in many systems of interest.

In recent years, a number of different approaches and methods have been proposed in order to include dispersion in 
semi-local DFAs~\cite{Dion_2004,Becke_2007,Riley_2010,Grimme_2010,Tkatchenko_2009a} (see Ref.~\onlinecite{Klimes_2012} for a comprehensive review). 
Among these, the pairwise-additive van der Waals (vdW) methods, which rely on a summation over inter-fragment $C_6/R^6$ terms, provide a 
simple and computationally efficient approximation for the long-range correlation energy~\cite{Klimes_2012} that is often capable of reliably 
describing certain classes of molecular systems. However, in many instances both quantitative and qualitative failures can occur, as was found 
in supramolecular systems,~\cite{Ambrosetti_2013} molecular crystals,~\cite{Roza_2012,Marom_2013,Reilly_2013a} and layered materials.~\cite{Nieminen-layered-2012} 
For example, pairwise methods are unable to correctly account for the experimentally well-known relative stabilities of the $\alpha$, $\beta$ and $\gamma$ polymorphs 
of the glycine molecular crystal,~\cite{Marom_2013} a shortcoming that is primarily due to the inherently many-body nature of long-range correlation and dispersion. 
While the pairwise contributions are sizeable, there are substantial higher-order energy and screening contributions, all of which must be considered in a collective 
many-body framework. In fact, the importance of many-body effects in the interatomic dispersion energy has been recently demonstrated in a number of 
studies.~\cite{Ambrosetti_2013,Reilly_2013a,Risthaus_2013,DiStasio_2012,Tkatchenko_2012a,Marom_2013}

The importance of many-body contributions to the correlation energy can be understood by making use of the random-phase approximation (RPA). 
When the correct ``bare'' susceptibilities are provided as input, RPA yields an accurate treatment of the long-range correlation energy, by virtue of the fact that RPA 
naturally accounts for higher order (beyond two-body) effects and electrodynamic response screening. Indeed, recent RPA implementations based on the adiabatic 
connection fluctuation-dissipation theorem (ACFDT) and Kohn-Sham orbitals yield an improved description of dispersion-bound systems with respect to 
semi-local DFAs.~\cite{Ren_2012,Furche-RPA-review,Dobson_2012} However, the steep computational cost associated with RPA@DFA calculations makes them impractical 
for most systems of interest.

RPA can also yield useful insight into the development of simpler and more efficient approaches for modeling many-body interactions. The infinite series resulting from a perturbative expansion of the RPA correlation energy is slowly converging, and the poor performance of low-order truncations~\cite{Ambrosetti_2013,Lu_2010} is clearly indicative of the limitations of low-order perturbative approaches with respect to the infinite-order many-body treatment. In this regard, it was observed that the simple inclusion of the Axilrod-Teller-Muto~\cite{Axilrod_1943} three-body term, also considered by others in the field,~\cite{Grimme_2010,Roza_2013} does not necessarily imply an improved accuracy with respect to a pairwise vdW expression.~\cite{Ambrosetti_2013}

Very recently, an alternative approach has been developed in order to compute the long-range many-body dispersion (MBD) energy in an efficient manner.~\cite{Tkatchenko_2012a,Silvestrelli_2013} Furthermore, an analytic proof has been presented demonstrating the equivalence between the MBD energy expression and the RPA correlation energy for a system of localized quantum harmonic oscillators coupled in the dipole approximation.~\cite{Tkatchenko_2013a} The MBD approach avoids the explicit use of single-electron orbitals, allowing for a favorable $N^3$ scaling (where $N$ is the number of atoms) and a negligible computational cost with respect to a self-consistent DFT calculation. This is made possible in the dipole approximation and for finite-gap systems, by mapping the ``bare'' susceptibility of the system onto that of a set of localized response functions. The MBD method~\cite{Tkatchenko_2012a} has proven to be very accurate for a wide variety of molecular and solid-state systems.~\cite{DiStasio_2012,Tkatchenko_2012,Marom_2013,Reilly_2013a}

In the present article we analyze in detail the basic concepts underlying the MBD method, and carefully consider the problem of polarizability screening and separating correlation into short-range and long-range contributions.
Here we introduce a revised MBD method (termed MBD@rsSCS), which employs range-separation (rs) of the self-consistent screening (SCS) of polarizabilities and the calculation of the long-range correlation energy. This enables us to separate the inherently short-range contribution to the correlation calculated with semi-local DFT from the long-range contribution calculated with MBD. The MBD@rsSCS method should be universally applicable to finite-gap systems, regardless of their size and geometry.

In the following sections (Sections~\ref{Ecor}--\ref{MBD_theory}) we will consider the derivation of the long-range correlation energy within the MBD framework before applying the MBD@rsSCS and related methods to a wide range of benchmark datasets in Section~\ref{app}, demonstrating the broad applicability and improved performance of the MBD@rsSCS method for a wide variety of systems. 

\section{The Correlation Energy in Density-Functional Approximations \label{Ecor}}

Common semi-local and hybrid xc functionals only provide a short-range description of the correlation energy. This follows from their integral definition, commonly given in terms of the density $n(\mathbf{r})$ and the correlation energy per particle $\epsilon_\text{c}$: 
\begin{equation}
E_\text{c}=\int \text{d}\mathbf{r} \, n(\mathbf{r}) \, \epsilon_\text{c}(n(\mathbf{r}),\nabla n(\mathbf{r})) \,.
\end{equation}
In absence of a truly non-local $\epsilon_\text{c}$,~\cite{Andersson_1996} the decay of the correlation energy contribution is directly controlled by that of $n(\mathbf{r})$. Hence, correlation effects in semi-local functionals will typically vanish exponentially with the distance between fragments with non-homogeneous electron densities. Therefore, the interaction energy between rare-gas atoms, for example, has an exponential decay, whereas a more long-ranged $\sim 1/R^6$ decay is expected.~\cite{Klimes_2012} 

An alternative and more powerful approach for calculating the correlation energy is to use the ACFDT formulation, wherein the correlation energy is expressed in terms of the density-density response function of the system $\chi(\mathbf{r},\mathbf{r'},i\omega)$, which describes the density response at $\mathbf{r}'$ induced by a perturbation at $\mathbf{r}$. The ACFDT formula employs the bare ($\chi_0$) response function of the non-interacting DFT system and the ``dressed'' ($\chi_{\lambda}$) response function of the interacting system, in which the Coulomb interaction $v$ is scaled by the adiabatic connection parameter $\lambda$:~\cite{Nguyen_2009} 
\begin{equation}
E_\text{c}= -\frac{1}{2\pi} \int_{0}^{\infty}\text{d}\omega \int_0^{1}\text{d}\lambda \, {\rm Tr}[\chi_{\lambda} v - \chi_0 v] \,.
\label{ACFDT}
\end{equation}

While $\chi_0$ is known analytically (as we shall discuss below in Section~\ref{resp}), the computation of the correlation energy from \eqnref{ACFDT} in practice requires the introduction of a model for $\chi_{\lambda}$. In this regard, one of the most popular approaches is the random-phase approximation (RPA), which corresponds to a summation of ring diagrams up to infinite order, neglecting effects in the interacting response function which derive from the xc kernel. Hence, the RPA correlation energy can be computed from \eqnref{ACFDT} by approximating $\chi_{\lambda}$ with the RPA interacting response function: 
\begin{equation}
\label{chi_lambda}
\chi_{\lambda}^\text{RPA}=\frac{\chi_0}{1 - \chi_0 \lambda v} \,.
\end{equation}
The ACFDT-RPA expression is commonly evaluated by computing $\chi_0$ using Kohn-Sham (KS) or Hartree-Fock (HF) electronic orbitals. While capable of describing different types of bonding, including ionic, covalent, and metallic bonds, RPA performs best at long range, as systematic underbinding is found for both molecules and solids,~\cite{Ren_2011a} which is attributed to the difficulty in accurately describing short-range correlation effects. Inclusion of renormalized exchange diagrams as in the SOSEX method,~\cite{Paier_2010} may yield improvements for ACFDT-RPA, but clearly this will further increase the already large computational cost of RPA based on KS or HF orbitals. 

A promising solution to this problem is offered by the so-called range-separation technique.~\cite{Toulouse_2004} In this framework, the electron-electron Coulomb interaction $v$ is split into short-range and long-range contributions: 
\begin{equation}
v=v_{\rm{SR}}+v_{\rm{LR}} \, .
\end{equation}
The correlation arising from the short-range part of the interaction $v_\text{SR}$ is efficiently treated using local or semi-local functionals, while the long-range part can be evaluated through traditional wavefunction methods, such as configuration interaction~\cite{Toulouse_2004} or RPA.~\cite{Toulouse_2010} The range-separation approach is therefore particularly convenient, as it allows for the application of different techniques that can be adapted for correctly describing both the short- and long-range correlation energy.

While the RPA correlation energy is normally computed using response functions calculated from KS or HF orbitals, the use of a model response function is equally possible. 
In this work we will exploit the frameworks of both RPA and range separation to compute the long-range electron correlation energy using a model system based on atom-centered response functions. As will be shown below, a judicious choice of model system and corresponding response functions can lead to a substantial reduction of the computational cost, while still yielding an accurate and physically motivated description of long-range correlation effects in non-metallic systems. An appropriate model system and corresponding response functions will be introduced in the next section. While we will employ a model system herein to compute the long-range correlation energy, we will only make use of quantities derived from the external potential $V_\text{ext}$ and the electron density $n(\mathbf{r})$. Thus, the long-range correlation energy will be consistent with HK picture of DFT, \emph{i.e.}, the long-range correlation energy will be a \emph{functional} of the density and external potential: $E_\text{c}^{\text{LR}}=E_\text{c}^{\text{LR}} [n(\mathbf{r}),V_\text{ext}]$.

\section{Model Response Functions for Free Atoms and Atoms in Matter \label{resp}}

As discussed above, a key quantity to be considered in the study of the electron correlation energy and response properties of a given molecular system is the (density-density) response function, $\chi(\mathbf{r},\mathbf{r}',i\omega)$. From the knowledge of $\chi(\mathbf{r},\mathbf{r}',i\omega)$, it is possible to obtain optical excitation spectra, polarizability, and higher multipolar susceptibilities of the system. Moreover, the ACFD theorem in \eqnref{ACFDT} provides an integral formulation of the correlation energy, given in terms of the bare and dressed response functions, and the Coulomb interaction. The ability to reliably compute the response function or susceptibilities of a system is thus a fundamental step towards an accurate and physically sound description of the correlation energy.

The response function $\chi_\lambda$ of the interacting system is typically written in terms of the bare susceptibility [\eqnref{chi_lambda}]. Therefore, we will first concentrate on $\chi_0$. Mathematically, $\chi_0$ can be determined from the complete set of eigenfunctions obtained from a mean-field electronic-structure calculation via the Adler-Wiser formalism:~\cite{Adler:1962,Wiser:1963}
\begin{equation}
\chi_0(\mathbf{r},\mathbf{r'},i\omega)= \sum_{ij}(f_i-f_j) \frac{\phi_i^*({\bf r})\phi_i({\bf r}')\phi_j^*({\bf r}')\phi_j({\bf r})}{\epsilon_i-\epsilon_j+i\omega} \,,
\label{chizero}
\end{equation}
where $\phi_i$ is the $i^\text{th}$ eigenorbital with corresponding eigenenergy $\epsilon_i$, and $f_i$ is the Fermi occupation number. However, the evaluation and the subsequent integration of the bare response function from electronic single-particle orbitals is computationally very expensive.

For a more efficient treatment of the response, we assume now that our system, consisting of molecules or condensed matter, has a finite electronic gap (between the highest-occupied and lowest-unoccupied molecular orbitals) and hence can be divided into effective atomic fragments. Under this assumption, we initially map the full nucleo--electronic system onto a set of localized atomic response functions. Since we are only interested in long-range effects, the atomic response is conveniently expressed in terms of multipole-multipole susceptibilities. The low-lying valence electron excitations, responsible for the long-range correlation effects, are then effectively described through a set of quasi-particles that are tailored in such a way as to reproduce the relevant multipole-multipole susceptibilities.

Given their simplicity and convenient analytical properties, we will represent the atomic response functions by a set of quantum harmonic oscillators (QHOs). Notably, only a single QHO per atom is needed to exactly reproduce the static dipole polarizabilities and homoatomic $C_6$ coefficients of isolated atoms.~\cite{Tkatchenko_2009a} This is readily shown from \eqnref{chizero}: by making use of the dipole-selection rules, the dipole polarizability tensor of an isotropic QHO having mass $m$, charge $Z$, and characteristic frequency $\tilde{\omega}$ can be written as:
\begin{equation}
\bm{\alpha}(i\omega)_{ln}=\delta_{ln}\frac{Z^2}{m(\tilde{\omega}^2+\omega^2)}\,,
\label{aqho}
\end{equation}
where the indices $l$ and $n$ indicate the Cartesian components of the tensor. Two conditions on the QHO parameters are then applied to obtain the exact value of $\bm{\alpha}(0)_{ii}$ (the isotropic static dipole polarizability) and  $\tilde{\omega}$. In the more general case of an anisotropic non-diagonal atomic dipole polarizability, a single SO(3)-rotated anisotropic QHO is still sufficient to reproduce the full static dipole polarizability tensor, in the absence of time-reversal symmetry breaking (non-imaginary eigenstates).

Analogous considerations hold for the higher multipolar response functions, which can again be expressed in terms of QHOs. Naturally, a larger number of QHOs will be necessary in order to exactly capture all the desired susceptibilities. Nonetheless, it was recently observed that a single isotropic QHO is sufficient for accurately reproduce dipole, quadrupole, and octupole response functions at the same time in noble gases, alkali atoms, and small molecules.~\cite{Jones_2013a}

In the present work we will make use of the dipole approximation to the Coulomb interaction $v$. Hence, from \eqnref{ACFDT}, only the dipolar component of the full response function is required. We remark that the full Coulomb potential can also be utilized to describe the coupling between atomic response functions, albeit at the expense of a much increased computational cost.~\cite{Jones_2013a} 

Among the different possible strategies for modeling the atomic dipole polarizability,~\cite{Becke_2007,Silvestrelli_2008,Andersson_1996,Vydrov_2010a} the Tkatchenko-Scheffler (TS)~\cite{Tkatchenko_2009a} method will be our method of choice in this work. This is particularly suitable as the TS approach is compatible with a single isotropic QHO per atom, and provides very accurate first-principles polarizabilities for both free atoms and small molecules. In particular, the free-atom polarizabilities and the corresponding homoatomic $C_6$ coefficients in the TS framework are derived from time-dependent density-functional theory (TDDFT) reference data,~\cite{Chu_2004} with an accuracy of $\sim$3\%. Similar accuracy has also been found for the corresponding heteronuclear $C_6$ coefficients. To account for the short-range exchange and correlation effects arising from the molecular environment, these free-atom polarizabilities are proportionally rescaled according to the atomic Hirshfeld volumes, computed from a given DFA electronic density. The resulting molecular $C_6$ coefficients show a mean absolute relative error (MARE) of $5.5\%$ for a database of 1225 small molecular dimers.~\cite{Tkatchenko_2009a}

As visible from the aforementioned performance for small molecular dimers, hybridization and short-range overlap effects are accurately treated through the TS procedure. In fact, the TS polarizabilities account for semi-local xc effects deriving from the DFT electron density.  However, while the Hirshfeld volume of a given atom is sensitive to the density modifications induced by neighboring atoms, the effects arising from more distant atoms decays exponentially with the distance, instead of the well-known electrostatic power law decay.

To explicitly account for the dipole interaction occurring between atomic response functions, we consider two charged atom-centered QHOs $i$ and $j$, which are located at $\mathbf{r}_i$ and $\mathbf{r}_j$, and separated by a distance $r_{ij}=|\mathbf{r}_i-\mathbf{r}_j|$. The Coulomb potential due to to these spherical Gaussian charge distributions (corresponding to the QHOs) is:~\cite{Tkatchenko_2012a,Tkatchenko_2013a}
\begin{equation}
v_{\rm{GG}}(r_{ij})=\frac{\text{erf}\left(r_{ij}/\sigma_{ij}\right)}{r_{ij}} \,.
\end{equation}
Here $\sigma_{ij}=\sqrt{\sigma_i^2+\sigma_j^2}$, with $\sigma_i=(\sqrt{2/\pi}\alpha_i/3)^{1/3}$ representing the Gaussian width of the $i^\text{th}$ QHO, which depends on its corresponding bare frequency-dependent polarizability. As we are operating within the dipole approximation, the coupling among QHOs is described solely by the dipole-dipole tensor derived from $v_\text{GG}$:
\begin{equation}
\mathbf{T}_{{\rm GG},ij}^{lm}= \partial_{r_i^l} \partial_{r_j^m} v_{\rm{GG}}(r_{ij}) \,,
\label{ggtensor}
\end{equation}
where $r_i^l$ indicates the $l^\text{th}$ Cartesian component of $\mathbf{r}_i$.

The long-range screened dipole polarizability is then computed through the discrete self-consistent screening (SCS) equation,~\cite{Tkatchenko_2012a}
\begin{equation}
\bm{\alpha}^{\rm{SCS}}(i\omega)=\bm{\alpha}^{0}(i\omega)-\bm{\alpha}^{0} \mathbf{T}_{\rm{GG}}\bm{\alpha}^{\rm{SCS}}(i\omega) \,,
\label{scsEQ}
\end{equation}
where $\bm{\alpha}^0_{lm}=\delta_{lm}\alpha^{\rm{TS}}_l$ is the diagonal bare polarizability matrix and $\bm{\alpha}^{\rm{SCS}}$ is the corresponding self-consistently screened polarizability matrix.

The intermolecular $C_6$ coefficients derived from the resulting SCS polarizabilities show a MARE of $6.3\%$ for the 1225 molecular dimer database discussed above. 
In addition, the anisotropy in the molecular polarizability is significantly improved by SCS compared to TS calculations that yield essentially isotropic
molecular polarizabilities. For more complex extended systems, calculations on Si clusters of increasing size and bulk Si yielded agreement within $\sim$8\% of $C_6$ coefficients derived from TDDFT and experiment, whereas for TS alone deviations up to $\sim$70\% were observed for the larger clusters.~\cite{Tkatchenko_2012a}

\section{The Long-Range Correlation Energy in the MBD Method \label{MBD_theory}}

In the preceding section, we introduced model response functions for computing the long-range correlation energy of finite-gap molecular systems. Here we outline the steps necessary to calculate an accurate long-range correlation energy utilizing these response functions, in a method we term MBD@rsSCS. 

\begin{figure}
\includegraphics[scale=0.3]{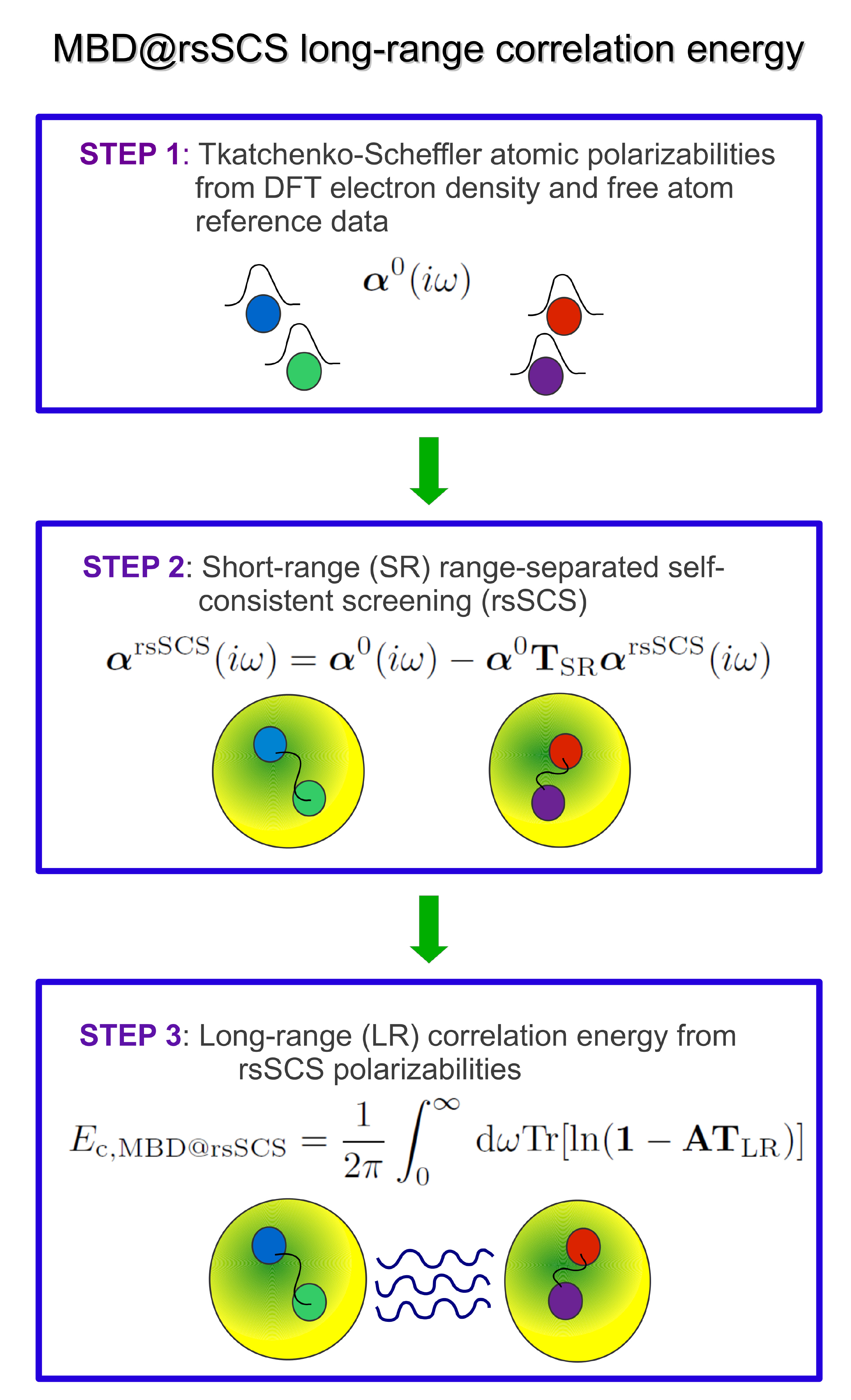}
\caption{Schematic diagram illustrating the computation of the MBD@rsSCS long-range correlation energy.
\label{fig1}}
\end{figure}

In particular, there are three key steps, as shown in \figref{fig1}:
\begin{enumerate}
\item Evaluation of the ``bare'' TS atomic polarizabilities from free-atom reference data and the DFT electron density, which accounts for short-range hybridization effects in the atomic polarizabilities.

\item Calculation of the self-consistently screened polarizabilities using range-separated self-consistent screening (rsSCS) of the TS atomic polarizabilities.

\item Computation of the many-body long-range correlation energy starting from the rsSCS input polarizabilities.
\end{enumerate}

To understand the benefit of the range separation between steps two and three, we begin by discussing the long-range correlation energy that arises from this model. Thereafter, we will discuss the concept of short-range polarizability screening, which follows naturally from the range separation of the coupling between QHOs.

\subsection{The Long-Range Correlation Energy from Coupled Atomic Response Functions}

Starting from the ACFDT formula [\eqnref{ACFDT}], the key ingredients contributing to the correlation energy are the response function $\chi$ and the interaction $v$. The single QHO approximation to the atomic response function has already been introduced and discussed above in Section~\ref{resp}. Here, we will focus on the Coulomb interaction.

In line with discussion of Section~\ref{Ecor} we begin by range-separating $v$ into $v_\text{SR}$ and $v_\text{LR}$. As the short- and long-wavelength excitations of the system will be typically associated with different energy scales, we neglect their explicit coupling as a first approximation, and express the total correlation energy as
\begin{equation}
E_\text{c} \simeq E_\text{c}^{\rm{SR}}+E_\text{c}^{\rm{LR}} \,,
\end{equation}
with the two terms stemming from $v_{\rm{SR}}$ and $v_{\rm{LR}}$, respectively.

Semi-local DFT represents an efficient and balanced choice for computing short-range correlation energy contributions. Therefore, we take $E_\text{c}^{\rm{SR}}$ to be the semi-local DFT correlation energy. We point out that at this point no explicit range separation of the electron-electron interaction is introduced in DFT. The desired separation is rather recovered from the implicit short-range character of the semi-local DFT correlation. A rigorous range separation of the correlation functional is also possible,~\cite{Toulouse_2004} and is currently under investigation.

In the present approach, the long-range correlation energy $E_\text{c}^\text{LR}$ can be formally defined through \eqnref{ACFDT} by substituting $v$ with $v_{\rm{LR}}$. This separate treatment of the long-range correlation energy contribution permits a multipole expansion of $v_{\rm{LR}}$ that can be restricted to lower orders. Clearly, at the large distances of interest here, the dipole component will be dominant compared to higher multipoles, which we will therefore neglect. As previously mentioned in Section~\ref{resp}, only the dipole component of the full response function will contribute to $E_\text{c}^{\rm{LR}}$ within the dipole approximation to $v_{\rm{LR}}$. Hence, we will again make use of the aforementioned single-QHO parametrization of the atomic response, as such a response function can reproduce static polarizabilities and homoatomic $C_6$ coefficients exactly [{\emph{cf.}, \eqnref{aqho}]. 


With respect to standard ACFDT-RPA implementations, in which $\chi_0$ is computed from electronic single-particle orbitals, the introduction of the MBD approach based on atomic response functions permits a considerable reduction in the degrees of freedom (\textit{i.e.}, to a single QHO per atom). As a result, in the specific case of dipole-coupled QHOs, \eqnref{ACFDT} can be recast into a very efficient discrete matrix formulation. This is accomplished after analytical integration over spatial variables by substituting $\chi_0 v$ with $\mathbf{AT}$.~\cite{Tkatchenko_2013a} Here $\mathbf{A}$ is a diagonal $3N \times 3N$ matrix, which is defined as $\mathbf{A}_{lm}=-\delta_{lm}\alpha_l(i\omega)$ in the case of isotropic QHOs, and $\mathbf{T}$ is now the dipole-dipole interaction tensor. The scalar $\alpha_l$ corresponds to the $l^\text{th}$ (isotropic) atomic polarizability computed using a chosen response function approximation (RFA). Extension to anisotropic polarizabilities is equally possible through the use of a block-diagonal $\mathbf{A}$ matrix.

Making use of this matrix formulation and considering the fact that ${\rm Tr}\lbrack \mathbf{A}\mathbf{T}\rbrack=0$ due to the absence of intra-oscillator interactions, the ACFDT-RPA correlation energy for the MBD model becomes:
\begin{equation}
E_{\rm{c,MBD}}=\frac{1}{2\pi}\int_0^{\infty} \, \text{d}\omega  {\rm Tr}\lbrack {\rm ln}(\mathbf{1}-\mathbf{AT})\rbrack \,.
\label{eMBD}
\end{equation}
In order for this energy expression to provide an accurate approximation to $E_c^{\rm{LR}}$, a range-separation of the interaction should be introduced in $\mathbf{T}$. Notably, if the tensor $\mathbf{T}$ is frequency independent, the same correlation energy as in \eqnref{eMBD} can be exactly computed by diagonalizing the MBD Hamiltonian (see Appendix~\ref{AppendA}).~\cite{Tkatchenko_2013a} This property makes the evaluation of the MBD correlation energy highly efficient, as the overall computational cost scales as $N^3$ (with $N$ being the number of atoms), and hence remains negligible with respect to the underlying DFT calculation.

We also emphasize that depending on the definition of the $\mathbf{T}$ tensor and the RFA used, different procedures for the computation of $E_{\rm{c,MBD}}$ can be defined. This is in direct analogy with standard ACFDT-RPA implementations, wherein different choices of the underlying orbitals might lead to different bare response functions and thus different correlation energies. Accordingly, our notation for the method employed will be MBD@RFA, similar to that used for RPA calculations, \emph{e.g.}, RPA@PBE.~\cite{Ren_2011a} 

In the original MBD implementation,~\cite{Tkatchenko_2012a} which we now refer to as MBD@SCS, the correlation energy was computed from the full (not range-separated) self-consistently screened TS polarizabilities. In this regard, we note that the full SCS equation \eqnref{scsEQ} provides RPA screening of the localized atomic polarizabilities. As such, long-range screening is also present in the RPA expressions discussed above. As a consequence, the MBD@SCS method effectively introduces a double RPA polarizability screening in the long range. The newly developed MBD@rsSCS method, which we will introduce in the following subsection, amends this aspect by providing single RPA screening at long distances.

\subsection{Range-Separation of the Interaction}

As shown in Section~\ref{resp}, the dipole-dipole interaction $\mathbf{T}_{\rm{GG}}$ that occurs between charged QHOs is derived from the Coulomb potential resulting from the corresponding Gaussian charge distributions [\eqnref{ggtensor}].~\cite{Tkatchenko_2012a,Tkatchenko_2013a} This interaction intrinsically contains a short-range attenuation due to the QHO charge overlap and is therefore well-suited for the treatment of the polarizability screening. The short-range behavior of $\mathbf{T}_{\rm{GG}}$, however, is not adequate for the computation of $E_\text{c}^{\rm{LR}}$, as in practice we have observed that a steeper attenuation of the interaction is required for coupling the DFT short-range correlation energy with the respective long-range component.

A potential solution to this problem is obtained by range-separating $\mathbf{T}_{\rm{GG}}$ into short- and long-range components, indicated by $\mathbf{T}_{\rm{SR}}$ and $\mathbf{T}_{\rm{LR}}$, respectively. We define the short-ranged $\mathbf{T}_{\rm{SR}}$ as:
\begin{equation}
\mathbf{T}_{{\rm SR},ij}^{lm}=(1-f(r_{ij}))\mathbf{T}_{\text{GG},ij}^{lm} \,,
\end{equation}
where $r_{ij}$ is the distance between atom $i$ and atom $j$, and $f(r_{ij})$ is the well-known Fermi-type damping function, widely employed in previous work:~\cite{WuYang-2002,Grimme2006,Tkatchenko_2009a,Silvestrelli_2008,Tkatchenko_2009a}
\begin{equation}
f(r_{ij})= \frac{1}{1+{\rm exp}[-a(r_{ij}/S_\text{vdW}-1)]}\,.
\label{fdamp}
\end{equation}
Here the $a$ parameter is fixed to a value of $a=6$, which leads to smooth damping behavior and substantially differs from the value typically used in pairwise vdW approaches in which $a \approx 20$. This is by no means surprising, as $f(r_{ij})$ directly acts on the dipole--dipole interaction here, which seamlessly controls the correlation energy at all orders. $S_\text{vdW}$ is defined as $\beta(R_{\rm{vdW}}^i+R_{\rm{vdW}}^j)$, where $R_{\rm{vdW}}^i$ is the vdW radius of the $i^\text{th}$ atom, and the parameter $\beta$ is fitted once per xc functional by minimizing energy deviations with respect to highly accurate reference data on a database of choice.

According to the range-separation technique, $\mathbf{T}_{\rm{LR}}$ is defined as $\mathbf{T}_{\rm{GG}}-\mathbf{T}_{\rm{SR}}$. The $\mathbf{T}_{\rm{GG}}$ tensor depends on the frequency of the electric field, while it is more computationally efficient to have a frequency independent interaction tensor $\mathbf{T}_{\rm{LR}}$. Therefore, we conveniently approximate $\mathbf{T}_{\rm{LR}}$ as:
\begin{equation}
\mathbf{T}_{{\rm LR},ij}^{ab} = f(r_{ij})\frac{-3r_{ij}^a r_{ij}^b+r^2_{ij}\delta_{ab}}{r_{ij}^5} \,,
\label{tensL}
\end{equation}
where $r_{ij}^a$ and $r_{ij}^b$ specify the $a$ and $b$ Cartesian components of $\mathbf{r}_{ij}$ and the vdW radii employed in $f(r_{ij})$ are consistent with the RFA used. The tensor is now frequency independent, and can be directly used in the MBD Hamiltonian (Appendix~\ref{AppendA}).

Within this range-separation approach, $\mathbf{T}_{\rm{SR}}$ is employed for the short-range self-consistent screening (SCS) of the TS polarizabilities: 
\begin{equation}
\bm{\alpha}^{\rm{rsSCS}}(i\omega)=\bm{\alpha}^{0}(i\omega)-\bm{\alpha}^{0} \mathbf{T}_{\rm{SR}}\bm{\alpha}^{\rm{rsSCS}}(i\omega) \,.
\end{equation}
All quantities are denoted here according to the notation introduced in Section~\ref{resp} above.

The resulting renormalized polarizabilities $\bm{\alpha}^{\rm{rsSCS}}$ are subsequently utilized as input parameters for the effective ACFDT-RPA computation of the long-range correlation energy $E_{\rm{c,MBD@rsSCS}}$. We remark that $E_{\rm{c,MBD@rsSCS}}$ is computed by making use of $\mathbf{T}_{\rm{LR}}$, \emph{i.e.}, the long-range component of the dipole-dipole interaction tensor, in order to avoid double counting of the DFT correlation energy and the polarizability screening.

Following from its definition, the MBD@rsSCS approach permits a ``soft'' short-range screening of the polarizability, providing at the same time a correct separation of the short- and long-range correlation energy. As will be shown later, this aspect is particularly relevant in condensed matter such as bulk semiconductors, where short-range screening is typically dominant over long-range effects. Significant contributions, however, are also found in the energetics of large supramolecular systems.

We emphasize here that no double counting of ring diagrams will be present in the MBD@rsSCS energy expression at long distances, preserving an effective RPA-like treatment of the many-body effects. This represents a major advantage with respect to the MBD@SCS method, especially in highly anisotropic systems, such as one- and two-dimensional nano-structures where long-range many-body effects become increasingly relevant.~\cite{Gobre_2013} The MBD@rsSCS method is also consistent with the overall DFT framework: the long-range correlation energy is computed by making use of the electron density and the atomic coordinates, and therefore can be expressed as $E_{\rm{c,MBD@rsSCS}}= E_{\rm{c,MBD@rsSCS}}[n(\mathbf{r}),V_\text{ext}]$.


\subsection{Interatomic Forces \label{forces}}

As well as providing a very efficient computation of the long-range correlation energy, the MBD model allows for the straightforward computation of interatomic forces for use in structural optimizations and molecular dynamics simulations.

To derive the interatomic forces, we will make use of the formulation given in \eqnref{eMBD} for $E_{\rm{c,MBD}}$. The interoscillator forces can now be obtained from the energy gradient as $\mathbf{F}=-\nabla E_{\rm{c,MBD}}$:
\begin{equation}
\mathbf{F}=\frac{1}{2\pi}\int_0^{\infty} \text{d}\omega \, {\rm Tr} \bigg{[} (\mathbf{1}-\mathbf{AT})^{-1} (\nabla \mathbf{AT} ) \bigg{]} \,.
\end{equation}
The differentiation of the $\mathbf{AT}$ matrix can be readily carried out if the dependence of the $\mathbf{A}$ matrix (and hence the polarizabilities) on the atomic positions is neglected. The subsequent numerical integration over imaginary frequency can be efficiently performed via Gauss-Legendre quadrature. Considering the dependence of the polarizability upon the atomic positions is also possible, and has recently been been studied in the context of self-consistent screening of the TS method.~\cite{Bucko_2013}

The present interatomic forces naturally account for the many-body effects present in the MBD energy. However, we stress that these only represent the long-range correlation contribution to the total interatomic forces. Due to the linearity of the gradient operator, the MBD forces can therefore be directly added to the forces originating from the underlying DFT calculation.

\section{Benchmarking and Applications \label{app}}

The MBD@rsSCS method is designed to accurately describe long-range correlation (and thus dispersion) in finite-gap systems, including at the same time a description of the short-range interactions from the underlying DFT computation of the electronic structure. To evaluate the performance of the DFT+MBD@rsSCS method, we will employ a wide variety of systems, representing different bonding types, symmetry, and physical extent, ranging from small-molecule dimers to supramolecular and solid-state systems. The DFT+MBD@rsSCS method will be assessed against the most accurate theoretical and experimental benchmark data available, allowing us to categorically ascertain the performance of the method as well as provide a broad overview of the importance of different dispersion contributions. In each application, the MBD@rsSCS method will be combined with the widely used PBE~\cite{Perdew_1996} semi-local functional and its corresponding hybrid functional PBE0,~\cite{Adamo_1999} which includes 25\% exact exchange. The comparison between the semi-local and hybrid functionals will also provide further insight into the role of exact exchange in these different systems.

The benchmarks utilized herein for small molecules are the S22 and S66$\times$8 databases of gas-phase dimers developed by Hobza and co-workers,~\cite{Jurecka_2006,Rezac_2011} for which ``gold standard'' coupled-cluster CCSD(T) binding energies are available. The performance for supramolecular and larger molecules will be assessed using the extrapolated experimental binding energies of the S12L database of Grimme.~\cite{Grimme_2012} Finally, the performance in the solid state will be assessed by comparison to the binding energy of bulk graphite and the X23 database of experimental molecular-crystals lattice energies.~\cite{Reilly_2013a,Reilly_2013c} All calculations were carried out using the all-electron FHI-aims code~\cite{Blum_2009}, employing the ``tier 2'' basis set, that essentially yields DFT binding energies at the basis set limit.

In addition to the MBD@rsSCS method, a second simplified many-body method will also be considered in this work, which follows from the ideas presented in Ref.~\cite{Tkatchenko_2013a}. This method consists of the direct use of unscreened TS polarizabilities as input parameters for MBD. The coupling among localized QHOs is again determined by the range-separated $\mathbf{T}_{{\rm LR}}$ interaction tensor, while the $\beta$ parameter, which controls the separation range, will be fixed independently of MBD@rsSCS. Although this approach lacks short-range polarizability screening, it is of particular interest as it allows for a straightforward implementation of interatomic forces (Section~\ref{forces}), and will be referred to as MBD@TS. 

\subsection{Parametrization of the MBD Method}

The range-separation parameter $\beta$ is determined by minimizing the mean absolute relative error (MARE) on the S66$\times$8 database.~\cite{Rezac_2011} As the S66$\times$8 database includes non-equilibrium geometries, this fitting procedure removes any potential bias towards equilibrium geometries. Performing the parametrization on small and medium-size molecules will further avoid the introduction of artifacts in large-scale systems. For such systems, changes in the cohesive energies will occur naturally due to many-body effects, and not due to \emph{ad hoc} modifications of the interaction.

The optimal values of $\beta$ are reported in \tabref{tabbeta} for both the MBD@rsSCS and MBD@TS methods with the PBE and PBE0 functionals. Interestingly, only slight variations are found between PBE and PBE0 for the value of $\beta$. This is certainly consistent with the fact that both PBE and PBE0 have the same correlation functional. The minor differences are therefore due to the fraction of exact exchange present in PBE0 and the corresponding improvement in its short-range treatment of xc effects.~\cite{Tkatchenko_2008a}

\begin{table}
\caption{Optimized values of the dimensionless $\beta$ parameter that controls the range-separation of the long-range correlation interaction.\label{tabbeta}}
\begin{ruledtabular}
\begin{tabular}{lcc}
           	& PBE 	& PBE0	   \\
\hline
MBD@rsSCS	& 0.83  & 0.85     \\  
MBD@TS	        & 0.81  & 0.83     \\  
\end{tabular}
\end{ruledtabular}
\end{table}

\subsection{Polarizability}

We begin by assessing the MBD method for its ability to determine $C_6$ dispersion coefficients. For small molecules there is only a marginal difference between the polarizabilities of MBD@rsSCS and MBD@TS. Considering a test set of 1225 molecular dimers,~\cite{Tkatchenko_2009a,Tkatchenko_2012a} the MARE in the $C_6$ coefficients is 6.2\% for MBD@TS and 7.1\% for MBD@rsSCS, while the pairwise TS method has a MARE of 5.5\%. However, we note here that the small size of systems in this database limits the importance of many-body contributions. In comparison, a recent assessment of the XDM method yielded a MARE of 10\% for a small-molecule database,~\cite{Roza_2013} while a number of vdW density functionals also yield larger deviations for small-molecule $C_6$ coefficients.~\cite{Vydrov_2010}

The effect of many-body contributions on the polarizability is far more pronounced in extended systems, in line with its importance for energetics. A clear example of this can be seen in the polarizabilities of bulk diamond and silicon. For diamond, the static isotropic TS polarizability is 10.9~a.u.\ per atom. In the absence of short-range screening MBD@TS gives a value within 1\% of this, as long-range screening effects cancel due to the high symmetry of the lattice. In contrast, including short-range effects as in MBD@rsSCS yields a polarizability of 7.2~a.u.\ per atom, in better agreement with the value of 5.6~a.u.\ extrapolated from the experimental dielectric constant \emph{via} the Clausius-Mossotti relation.~\cite{Bermejo_1982} The same is true for silicon, where the TS value of 33.6~a.u.\ is essentially unvaried with MBD@TS, but is reduced to 24.5~a.u.\ by MBD@rsSCS, in good agreement with a TDLDA benchmark value of 26.6~a.u.~\cite{Zhang_2011} The original MBD@SCS approach has also been shown to give good predictions for the dielectric properties of the molecular crystals of the acene molecules.~\cite{Schatschneider_2013}

\subsection{Small- and Medium-Sized Molecules}

We now turn to the assessment of the energetics of MBD, starting with a benchmark of the DFA+MBD@rsSCS and DFA+MBD@TS methods using the the S22~\cite{Jurecka_2006} and S66$\times$8~\cite{Rezac_2011} databases of Hobza and co-workers. These databases contain a variety of gas-phase dimers (22 and 66 in the S22 and S66, respectively), covering different geometrical structures and bonding types, with binding energies computed by means of coupled-cluster theory with single, double, and perturbative triple excitations [CCSD(T)]. In the  case of the S66$\times$8, eight different inter-fragment distances are considered, with the equilibrium configurations being referred to as the S66 database. This permits accurate benchmarking using more challenging non-equilibrium geometries. In addition, a subset consisting of the seven largest distances is also considered in the present work and denoted as S66$\times$7. With this database the many-body dispersion contribution in the long-range limit can be better assessed.

\begin{table*}
\caption{MARE (in \%) of different vdW-inclusive DFT methods with MAE (in kcal/mol) given in parenthesis.\label{tabsmall}}
\begin{ruledtabular}
\begin{tabular}{lrrrrrrr}
 	        & PBE+MBD@rsSCS	& PBE0+MBD@rsSCS& PBE+MBD@TS 	& PBE0+MBD@TS	& PBE+TS     	& PBE0+TS     \\
\hline
S22             &  8.9 (0.49)	& 8.5 (0.55)    &  8.6 (0.45)	&  8.7 (0.53)	&  9.1 (0.29)	&  7.3 (0.30) \\  
S66             &  9.0 (0.42)	& 8.1 (0.40)    & 10.0 (0.45)	&  8.9 (0.43)	& 12.0 (0.44)	&  9.7 (0.40) \\  
S66$\times$8    & 10.6 (0.32)	& 9.2 (0.30)    & 12.3 (0.36)	& 10.5 (0.33)	& 13.6 (0.37)	& 11.1 (0.33) \\  
S66$\times$7    &  7.9 (0.27)	& 6.4 (0.25)    & 10.3 (0.32)	&  8.3 (0.29)	& 13.0 (0.34)	& 10.2 (0.30) \\  
\end{tabular}
\end{ruledtabular}
\end{table*}

The mean absolute relative errors (MARE) and mean absolute errors (MAE) with respect to the CCSD(T) reference data are presented in \tabref{tabsmall}. In addition to the many-body schemes the values for the pairwise TS method are also given. Generally, all of the methods yield good performance with MAEs well below the 1~kcal/mol chemical accuracy threshold. It is also clear that for the larger S66 datasets PBE0 yields a noticeable improvement over PBE. The inclusion of many-body contributions also substantially improves over the pairwise TS method, which has a comparable MAE to Grimme's PBE-D3(BJ) for the S66$\times$8.~\cite{Goerigk_2011,Reilly_2013a} The inclusion of short-range screening in MBD@rsSCS also yields expected improvements in binding energies.

For the S22 the picture is less clear. First, TS performs very well for this dataset. One reason for this is the fact that the TS damping function was fitted to the S22,~\cite{Tkatchenko_2009a} making it optimal for this database. The majority of the error arises in complexes 11 to 15, which are stacked aromatic dimers. MBD@rsSCS yields systematic underbinding for these complexes and as MBD@TS systematically overbinds other systems it appears to perform better here. Accurate treatment of anisotropy in the molecular polarizabilities may play an important role in such small systems, for which an accurate treatment of anisotropy in the short-range interaction is crucial. In the stacked dimers much of the polarization will be in the plane of the molecules. When this is isotropized for input into the MBD Hamiltonian, the polarizability in the perpendicular direction will be overestimated, possibly leading to an over-screening of the stacking interaction. However, it is clear that for larger systems, and a broad benchmark like the S66, MBD@rsSCS represents a substantial improvement over the other methods. This will be further confirmed by application to more extended systems in the following subsections. 

\begin{figure}
\includegraphics[scale=0.32]{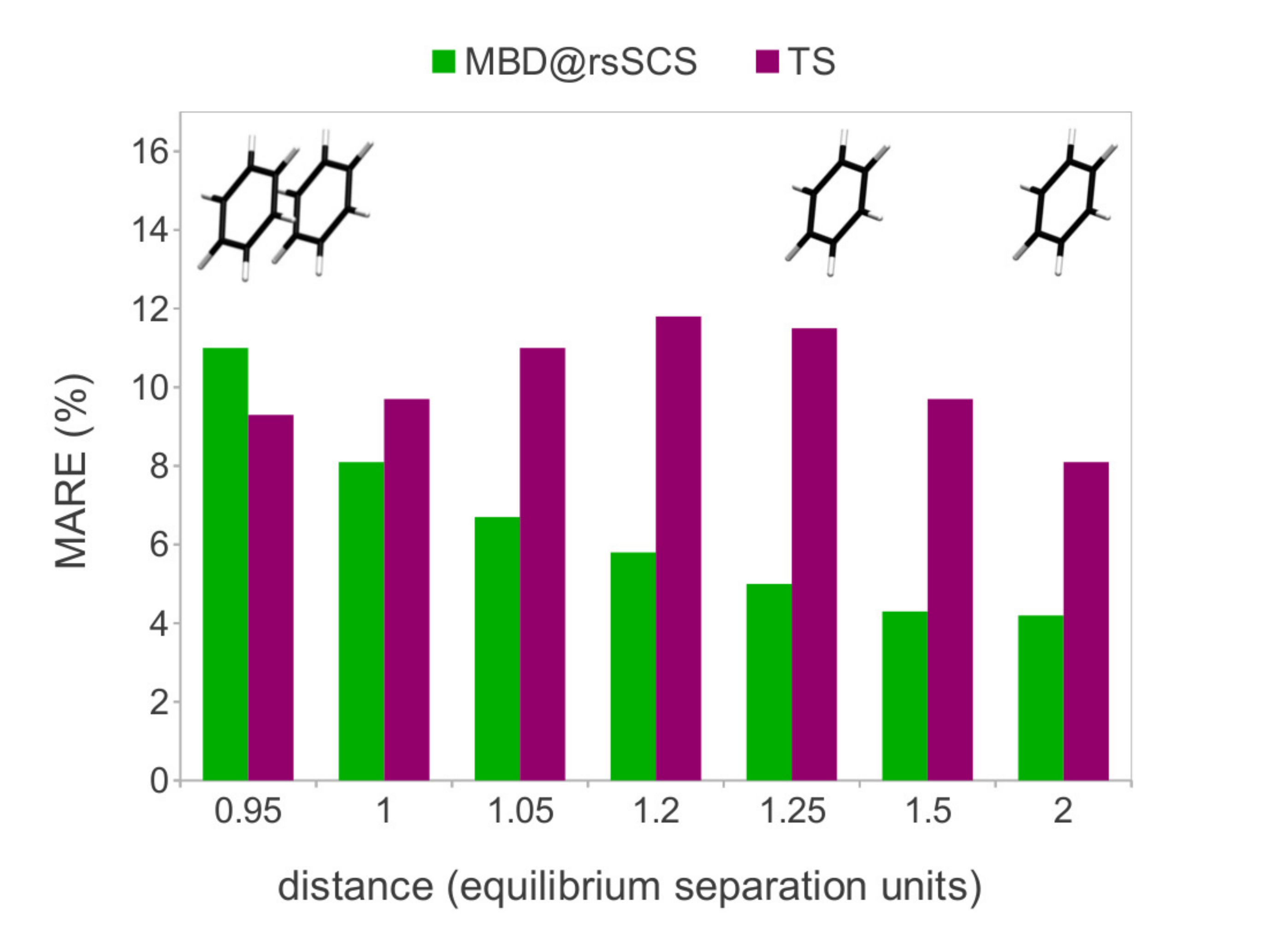}
\caption{The mean absolute relative error of PBE0+MBD@rsSCS and PBE0+TS (in \%) at each of the S66$\times$7 inter-fragment separations.
\label{fig2}}
\end{figure}

We now focus our attention on how the accuracy of the various methods changes as a function of the inter-fragment separation. The description of short-range contributions is expected to be intrinsically difficult for both MBD@rsSCS and MBD@TS as these theories consider only dipole-dipole contributions and were derived on the assumption that the interacting fragments were well separated. This is indeed the case, as a clear improvement in the accuracy is observed when omitting the shortest distance, as in the S66 and S66$\times$7 datasets. Higher multipole components of the Coulomb interaction become increasingly important at the short range, while a detailed description of the effects of overlap between electronic orbitals is not possible with a single isotropic dipole oscillator per atom.

We further investigate this aspect through a more detailed analysis of the error as a function of the inter-fragment distance, as shown in Table~\ref{tabdist} for PBE0+MBD@rsSCS. A monotonic decrease of both the MARE and MAE is found with increasing distances, while the deviation found at the shortest separations (0.9 times the equilibrium distance) reaches 29\% of the total energy. However, even here the MAE remains well below 1 kcal/mol, still maintaining chemical accuracy. The consistent improvement in the accuracy found with larger distances (See ~\figref{fig2}) provides a clear confirmation of the quality and robustness of the MBD model as a method for modeling long-range correlation effects.

\begin{table}
\caption{MARE (in \%) and MAE (in kcal/mol) of PBE0+MBD@rsSCS for each of the separations in the S66$\times$8 database. Separations are given as multiples of the equilibrium distance.\label{tabdist}}
\begin{ruledtabular}
\begin{tabular}{lcc}
Separation &MARE (\%) & MAE (kcal/mol) \\
\hline
0.90	& 29.0 	& 0.64	\\
0.95	& 11.0 	& 0.50	\\
1.00	& 8.1 	& 0.40	\\ 
1.05	& 6.7 	& 0.32	\\
1.10	& 5.8 	& 0.26	\\
1.25	& 5.0 	& 0.16	\\
1.50	& 4.3 	& 0.08	\\
2.00	& 4.2 	& 0.02	\\
\end{tabular}
\end{ruledtabular}
\end{table}

There are a number of pairwise dispersion methods that have been applied to the S66 database or parts thereof. These include the TS method (already discussed above), DFT-D3,~\cite{Grimme_2010,Goerigk_2011} and the non-local vdW-DF2~\cite{Lee_2010} and VV10~\cite{Vydrov_2010a} functionals. Typically their MAEs are within 0.2~kcal/mol of MBD@rsSCS, close to the limit of the accuracy and convergence of the reference data. Indeed, despite the inherent differences between the different methods, their performance for small molecules is generally good, implying that a pairwise description accounts for the majority (but not all) of the long-range correlation effects in these systems, in agreement with the work of Grimme.~\cite{Grimme_2010} While the role of many-body vdW interactions is limited in these systems, this will not be the case in general for larger molecules and condensed phases, as we shall demonstrate in the following subsections.

\subsection{Supramolecular Systems}

As a benchmark for larger and more complex intermolecular interactions, we employed the S12L dataset of Grimme.~\cite{Grimme_2012} This database is comprised of 12 supramolecular complexes, formed by six hosts combined with two guest monomers each.~\cite{Grimme_2012,Risthaus_2013} Due to the mixed nature of the non-covalent host-guest interactions, which includes hydrogen bonding, dispersion, $\pi$-$\pi$ stacking, and electrostatic (cation-dipolar) bonding, the S12L dataset can be regarded as representative of large host-guest systems. The reference binding energies were determined by extrapolation from experimental association free energies through the use of approximate solvation and entropic corrections.~\cite{Grimme_2012} These energies have an estimated average error of 2~kcal/mol.~\cite{Grimme_2012,Risthaus_2013}

Due to the larger size of these complexes, which contain between 86 and 177 atoms, and their relatively anisotropic geometries, we expect that many-body dispersion effects will play a much more significant role than for the smaller molecules in the preceding section. Indeed, recently it has been shown that the many-body contributions of MBD@rsSCS yield a substantial reduction of the host-guest binding interactions in all of the systems, typically on the order of 10\% of the energy.~\cite{Ambrosetti_2013} This amounts to PBE+MBD@rsSCS having a MAE of 1.6~kcal/mol (MARE 5.4\%),~\cite{Ambrosetti_2013} which is within the estimated uncertainty in the reference data.~\cite{Grimme_2012,Risthaus_2013} In contrast, PBE+TS has a MAE of 8.0~kcal/mol (MARE 25\%), corresponding to a systematic overestimation of the host-guest binding. From \tabref{tabs12l} we can see that PBE+MBD@TS also performs substantially better than PBE+TS. PBE+MBD@TS still overbinds due to the lack of proper short-range polarizability screening. As already observed and discussed elsewhere,~\cite{Ambrosetti_2013} the combination of a vdW term (either pairwise or many-body) with PBE always performs better than with the hybrid PBE0 functional, which is in contrast to the behavior seen for the smaller molecules of the S22 and S66 databases. This is possibly related to the larger \emph{contact area} between host and guest monomers, which implies substantial overlap effects, and might require a more sophisticated treatment of exchange, including for instance screening effects.~\cite{Heyd_2003,Paier_2010}

\begin{figure}
\includegraphics[scale=0.32]{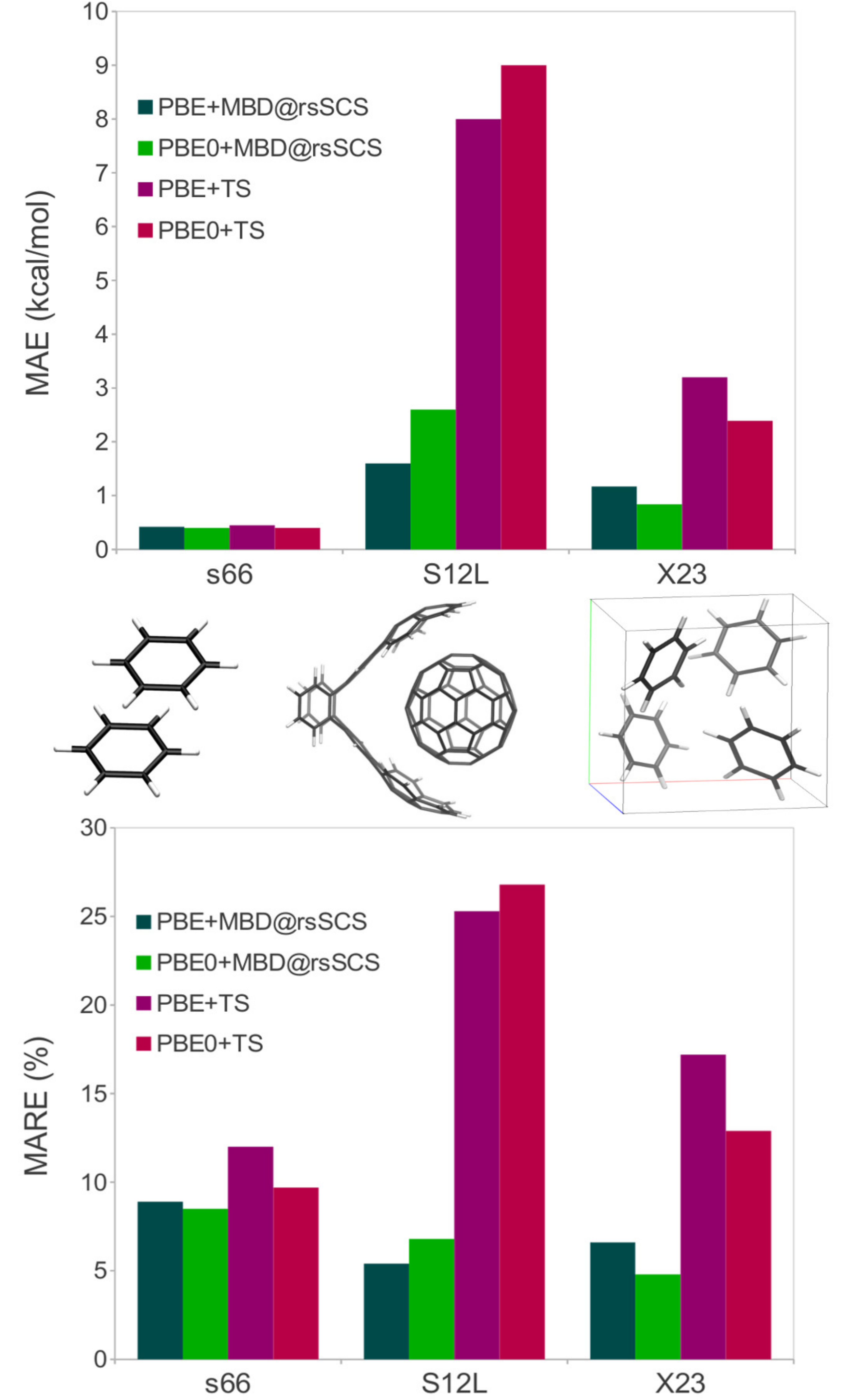}
\caption{MAE (upper panel in kcal/mol) and MARE (lower panel in \%) for the S66, S12L and X23 databases computed with MBD@rsSCS and TS combined with both the PBE and PBE0 functionals.\label{fig3}}
\end{figure}

\begin{table}
\caption{MARE (in \%) and MAE (in kcal/mol) in the binding energies of various vdW-inclusive DFT methods when compared to the experimental binding energies of the S12L database.~\cite{Grimme_2012} The MBD@rsSCS values are taken from Ref.~\cite{Ambrosetti_2013}, while the PBE-D3 values are from Ref.~\cite{Risthaus_2013}.\label{tabs12l}}
\begin{ruledtabular}
\begin{tabular}{lcc}
	       & MARE (\%) & MAE (kcal/mol) 	\\
\hline
PBE+MBD@rsSCS  &  5.4 	& 1.6	\\
PBE0+MBD@rsSCS &  6.8 	& 2.6	\\
PBE+MBD@TS     & 10.0 	& 3.3	\\
PBE0+MBD@TS    & 12.1 	& 4.5	\\
PBE+TS         & 25.3 	& 8.0	\\
PBE0+TS        & 26.8 	& 9.0	\\
PBE-D3         &  5.9 	& 2.1	\\
\end{tabular}
\end{ruledtabular}
\end{table}

A detailed analysis of the many-body contributions has shown that high-order contributions (typically up to sixth-order) are required to achieve reasonable convergence (better than 1~kcal/mol) with the full infinite-order MBD@rsSCS results.~\cite{Ambrosetti_2013} Considering a number of (effective) pairwise approaches to dispersion interactions,~\cite{Risthaus_2013,Ambrosetti_2013} the D3 method performs the best for these supramolecular systems, with PBE+D3 having a MAE of 2.1~kcal/mol, within 0.5~kcal/mol of the PBE+MBD@rsSCS value. It is important to note though that many-body contributions can be very system dependent.~\cite{Ambrosetti_2013} As the many-body contributions are very sensitive to the size, geometry, and electronic properties of a given system, it is difficult to \emph{a priori} assess the performance of more-empirical pairwise methods, which often have fixed or only locally dependent dispersion coefficients. In contrast, MBD@rsSCS is able to naturally ``adapt'' to the different systems through its seamless consideration of many-body effects, yielding higher accuracy and good transferability for very different types of systems (See also ~\figref{fig3}).

\subsection{Molecular Crystals}

Moving to even more extended systems we consider the lattice energy of molecular crystals. The X23 database consists of 23 molecular crystals,~\cite{Reilly_2013a,Reilly_2013c} and was derived in part from the database of Otero de la Roza and Johnson.~\cite{Roza_2012} The X23 database is comprised of 11 systems containing predominately dispersion interactions between molecules, 9 hydrogen-bonded systems, and 3 systems that feature both types of interactions. The experimental lattice energies were determined using experimental sublimation enthalpies with vibrational contributions from phonon calculations and experimental heat-capacity data, with an estimated uncertainty of 0.5--0.75~kcal/mol.~\cite{Reilly_2013c}

The MAEs and MAREs for various methods are presented in \tabref{tabx16}. As noted previously, the pairwise TS method performs poorly for molecular crystals, with a MAE of 2.39~kcal/mol when combined with PBE0.~\cite{Reilly_2013a,Reilly_2013c} Just as for supramolecular systems, TS overestimates cohesion due to the absence of long-range screening contributions. MBD@SCS yields substantial improvements with PBE0+MBD@SCS having an MAE of 0.94 kcal/mol,~\cite{Reilly_2013c} below the coveted chemical accuracy standard of 1 kcal/mol. Range-separation of the electrodynamic screening further improves the performance, with PBE0+MBD@rsSCS reaching a MAE of 0.84 kcal/mol. As for small molecules, including some exact exchange in the underlying density functional improves performance by a significant amount, although the role of many-body vdW contributions is much larger. This may in part be due to a better balance between the modeling of the crystal and the isolated molecules. Delocalization is inherent in the solid state, in particular for hydrogen-bonded systems, so that self-interaction errors are less of an issue. Using the hybrid functional therefore improves the description of the isolated molecules. In addition, as discussed previously,~\cite{Ambrosetti_2013} the ``contact area'' of these systems is reduced, \emph{i.e.}, there is only limited inter-fragment charge overlap, giving a rather clean separation between short- and long-range effects. 

MBD@TS performs markedly better than TS for X23, as was the case for the S12L. However, we can clearly see the importance of the short-range screening that is present in MBD@rsSCS, with the PBE0+MBD@TS MAE being 0.50~kcal/mol larger than PBE0+MBD@rsSCS. In the absence of this screening, MBD@TS overbinds the crystals. This is particularly visible in the highly polarizable adamantane crystal, where MBD@TS predicts a cohesive energy 4.5~kcal/mol larger than that of MBD@rsSCS. At short-range, \emph{i.e.}, within molecules and close fragments, screening of the system can result in strong polarization along chemical bonds,~\cite{Reilly_2013a} increasing the strength of the long-range coupling and screening of the dipole oscillators. This in turn can lead to stronger depolarization and therefore an overall greater reduction in relative binding energies.

For 21 of the X23 systems a comparison with pairwise methods can be made using the C21 data of Otero-de-Roza and Johnson.~\cite{Roza_2012,Reilly_2013c} For this dataset, PBE0+MBD@rsSCS achieves an MAE of 0.76~kcal/mol while PBE+MBD@rsSCS has an MAE of 1.07~kcal/mol. In contrast, both B86b+XDM~\cite{Roza_2012a,Becke_2007} and the vdW-DF2 functional~\cite{Lee_2010} give larger MAEs of 1.13~kcal/mol and 1.53~kcal/mol, respectively. PBE+MBD@rsSCS can be improved to yield chemical accuracy by inclusion of exact exchange (as in PBE0+MBD@rsSCS). However, systematically improving XDM or vdW-DF2 is difficult. B86a+XDM has a negative mean error of $-$0.83~kcal/mol, underbinding the crystals. Correctly adding exchange contributions or many-body dispersion effects would in fact worsen its performance, as their effect is to reduce lattice energies even further. Indeed, it has already been seen that adding three-body dispersion energy contributions yields a slightly poorer MAE.~\cite{Roza_2013} Adding higher-order contributions to vdW-DF2, in the form of triple or higher non-local integrals, would raise its computational cost dramatically. The ability to systematically improve the performance of MBD through inclusion of different ranges of screening and many-body contributions (\emph{e.g.}, @TS and @rsSCS) and combination with hybrid functionals is another particular strength of this approach.

\begin{table}
\caption{MARE (in \%) and MAE (in kcal/mol) for the X23 database calculated with different vdW-inclusive DFT schemes. The PBE+TS and PBE0+TS values are taken from Ref.~\cite{Reilly_2013a}.\label{tabx16}}
\begin{ruledtabular}
\begin{tabular}{lcc}
Method         & MARE (\%) 	& MAE (kcal/mol) 	\\
\hline
PBE+MBD@rsSCS	& 6.6  	& 1.18	\\
PBE0+MBD@rsSCS	& 4.8 	& 0.84	\\
PBE+MBD@TS   	& 10.0 	& 1.76	\\
PBE0+MBD@TS	& 7.7 	& 1.34	\\
PBE+TS	       & 17.2 	& 3.20	\\
PBE0+TS        & 12.9 	& 2.39  \\
\end{tabular}
\end{ruledtabular}
\end{table}

\subsection{Graphite Interlayer Binding}

As a final example of the performance of the MBD method we consider the interlayer binding energy of bulk graphite with AB stacking at the experimental interlayer equilibrium distance (3.35~\AA)~\cite{Chung_2002}. Given their highly anisotropic nature, graphite and graphene represent a particularly challenging test case for all dispersion methods. Although graphite does not have a finite gap, only a small fraction of the charge has metallic character. Hence, starting from atomic fragments still represents a reasonable approach to the computation of the long-range correlation energy in this system.

Recently, slow convergence of many-body effects has been observed in graphene,~\cite{Gobre_2013} due to the two-dimensional nature of individual sheets of graphene. This suggests that many-body effects will likely play a significant role in governing the binding energy in graphite. This also implies that a large supercell will be required to converge the long-range collective excitations that govern the correlation behavior. For this reason, the long-range correlation energy has been computed by modeling bulk graphite with an 11$\times$11$\times$7 supercell of the unit cell, which corresponds to 3388 atoms. Moreover, interactions with periodic replicas of atoms were considered up to a cut-off radius of 200~\AA .

\begin{table}
\caption{DFT-calculated interlayer binding energies (in meV per C atom) of bulk graphite at the experimental interlayer distance.\label{tabgraf}}
\begin{ruledtabular}
\begin{tabular}{l c}
Method & E$_\text{bind}$ (meV/C) \\
\hline
PBE+MBD@rsSCS   &    48     \\	
PBE0+MBD@rsSCS  &    50     \\
PBE+MBD@TS      &    56     \\
PBE0+MBD@TS     &    57     \\
\end{tabular}
\end{ruledtabular}
\end{table}

The interlayer binding of graphene with MBD is shown in \tabref{tabgraf}. MBD@SCS is not reported in \tabref{tabgraf} as the isotropization of long-range screened polarizabilities in the MBD Hamiltonian yields an overestimate of the cohesion of individual graphene sheets. As a consequence, no stability is predicted by MBD@SCS for superposed graphene layers. In contrast, MBD@rsSCS, thanks to its correct RPA treatment of the long-range interaction, avoids the problems related to highly anisotropic systems, and for PBE0+MBD@rsSCS yields an interlayer binding energy of 50 meV per C atom, in excellent agreement with an experimental value of 52$\pm$2~meV.~\cite{Zacharia_2004} Good agreement is also found with the RPA calculation of Dobson and co-workers,~\cite{Lebegue_2010} which predicts a binding energy of 48 meV. The difference between PBE0+MBD@rsSCS and PBE+MBD@rsSCS amounts to only 2 meV, \emph{i.e.}, within the experimental error bar. As expected from the results for other extended systems, MBD@TS gives slightly overestimated binding, while PBE+TS severely overestimates the cohesion with an energy of 87 meV.

For two-layer cohesion of graphite, values of 66 and 35 meV were reported for PBE-D2 and PBE-D3, respectively.~\cite{Grimme_2010}
While we can not directly compare with the bulk values calculated here, the PBE-D2 estimate presumably is likely too high, as in the case of PBE+TS. 
Notably, the PBE-D3 energy appears to be substantially lower. In part this change stems from a reduction of the pairwise component due to a 
renormalization of the interatomic $C_6$ coefficients, but a significant contribution is likely due to the inclusion of the three-body ATM energy term. 
This term has a repulsive character, and could lead to a substantial underestimation of the dispersion energy in the absence of counter-corrections 
coming from higher than three-body contributions.~\cite{Ambrosetti_2013}

\section{Discussion}

We now briefly summarize and discuss the application of the MBD methods to the various benchmark systems. For all but the smallest of the benchmark systems, MBD@rsSCS yields a systematic improvement in binding and cohesion, showing good applicability for systems ranging from small molecules to complex supramolecular and solid-state assemblies. The sole outlier in this regard are certain systems in the S22 database where underbinding is observed. However, this is likely due to subtle effects and contributions that are poorly described by approximate DFT functionals that do not accurately treat short-range anisotropic interactions. From the good performance of the MBD method for larger, more complex systems, it is clear that MBD accurately accounts for the most important long-range contributions. Comparison with pairwise methods, which neglect the collective many-body nature of dispersion interactions, clearly illustrates the importance of many-body contributions in more complex systems, with MBD@rsSCS reaching the uncertainty of the benchmark data for the supramolecular and molecular-crystal systems.

The range-separation approach to many-body dispersion also yields further insight into the role of short- and long-range correlation energy contributions. MBD@TS considers only long-range screening and collective contributions to the many-body dispersion energy, and consequently overestimates binding. In bulk Si, where long-range collective contributions largely cancel, neglecting short-range screening yields essentially the same polarizability as the underlying pairwise-additive TS method. This illustrates that an accurate determination of the long-range correlation energy can require accurate modeling of correlation effects at all ranges, due to their fundamentally collective origin. MBD@rsSCS allows one to capture the short-ranged screening contributions and their influence on the long-range correlation energy, while still employing the semi-local DFT correlation at short ranges. This range-separation approach also removes the spurious double counting of long-range screening effects present in MBD@SCS, and further improves on its performance.

In the present work we have employed both a semi-local functional (PBE)~\cite{Perdew_1996} and its hybrid variant (PBE0).~\cite{Adamo_1999} For small molecules and molecular crystals, the use of a hybrid functional yields systematic improvements in the performance of MBD@rsSCS. In the former case this is likely due to a better description of short-ranged correlation,~\cite{Tkatchenko_2008a} while in the latter case the hybrid functional likely gives a better description of the isolated molecules that the lattice energy is calculated relative to.~\cite{Reilly_2013a} However, PBE0 yields poorer performance for the supramolecular systems, possibly due to the greater overlap between fragments in these systems, which leads to different contributions than observed in simple dimers or molecular crystals. It is clear that hybrid functionals have a different role in modeling cohesion in the different benchmark systems--a role that apparently influences correlation in an indirect manner. It is evident that these hybrid functional contributions must be understood on a more fundamental level, and this is one focus of ongoing research. More rigorous range separation including that of the exchange, or even the use of screened exact-exchange functionals,~\cite{Heyd_2003,Paier_2010} may offer potential solutions. In this regard we also note that MBD@rsSCS need not be restricted to employing DFT for computing the short-range correlation. The MBD@rsSCS approach could be combined with other range-separated electronic-structure methods, or even more semi- or even fully empirical approaches. However, we stress that in combining MBD@rsSCS with DFT we have an approach that yields a robust and transferable description of a wide variety of systems.

As a final note, we emphasize that the computational cost of the MBD approach is largely negligible compared to the underlying DFT calculation, requiring only standard matrix operations such as diagonalization, on $3N\times 3N$ matrices. There is considerable scope for optimization and parallelization of these operations, as well as the potential for straightforward computation of interatomic forces (Section~\ref{forces}). As such, this approach will become suitable for performing \emph{ab initio} molecular dynamics simulations with an accurate modeling of long-range correlation energies. 

\section{Conclusions}

In this work we have developed a highly efficient and accurate method (MBD@rsSCS) for the calculation of the long-range correlation energy in finite-gap systems. This has been achieved in the dipole approximation by making use of a range-separated electron-electron interaction and localized atomic response functions. The range separation enables us to account for short-range correlation using a semi-local or hybrid functional, while long-range contributions are accounted for by an effective RPA model based on coupled atomic response functions. The range separation step only requires a single parameter, which is fitted to accurate quantum chemistry benchmark data.

Application of this approach to accurate benchmark datasets yields small mean absolute errors superior to pairwise models of dispersion, illustrating the importance of many-body contributions to correlation energies and the ability of MBD@rsSCS to seamlessly account for them. In general, the role of many-body contributions is to reduce cohesion by reducing dispersion interactions. A related method, MBD@TS, which contains only long-range screening contributions to the long-range correlation energy, yields poorer performance, highlighting the importance of short-range screening and the need for accurately balancing short-range and long-ranged contributions to the correlation energy.

Combining MBD@rsSCS with hybrid density functionals, specifically PBE0 in this work, reveals mixed performance for the benchmark systems. Whereas PBE0 leads to significant improvements for small molecular dimers and molecular crystals, the description of binding in supramolecular systems is worse. This points to the different role and contributions of exact exchange in the cohesion of different types of molecular systems. The application of more sophisticated treatments for exact exchange~\cite{Heyd_2003,Paier_2010} may lead to a more fundamental understanding of their role and more uniform overall performance.

The MBD@rsSCS method is both accurate and efficient, with a computational cost that is negligible with respect to the underlying DFT calculation. With the possibility of an efficient determination of interatomic forces, this method can be applied to a wide range of challenging and complex systems, from molecules to solids, with an accurate and physically motivated description of long-range correlation, even for the study of dynamic properties.

\appendix
\section{Direct Computation of the Long-Range Correlation Energy from the MBD Model Hamiltonian \label{AppendA}}
The computation of the many-body dispersion energy $E_\text{c}^{\rm{LR}}$ follows from a model of isotropic atom-centered QHOs, which are coupled through a range-separated dipole-dipole interaction. The corresponding MBD model Hamiltonian is defined as:~\cite{Tkatchenko_2012a,Bade_1957a}
\begin{equation}
\hat{H}_{\mbox{\scriptsize {MBD}}}=-\sum_{p=1}^N \frac{\nabla^2_{\bm{\xi}_p}}{2} +\sum_{p=1}^N \frac{\omega_p^2 \bm{\xi}_p^2}{2}+
\sum_{p>q}^N \omega_p \omega_q \sqrt{\alpha_p \alpha_q} \bm{\xi}_p \mathbf{T}_{pq} \bm{\xi}_q \,,
\label{hamilt}
\end{equation}
where $\bm{\xi}_p$ represents the mass-weighted QHO displacement from the $p^\text{th}$ atomic position, and $\alpha_p$ and $\omega_p$ represent the input isotropic dipole polarizability and characteristic frequency for the $p^\text{th}$ atom, respectively, \emph{e.g.}, as determined from the rsSCS procedure. The first two terms in the Hamiltonian represent the kinetic and the potential energy of the individual QHOs. The third is the two-body coupling among QHOs due to the dipole-dipole tensor $\mathbf{T}$. As discussed in the main text, there are a number of different sets of polarizabilities and frequencies that can be used to approximate the response of the system (\emph{i.e.}, TS, SCS, and rsSCS).

As the interoscillator potential energy is bilinear in the spacial displacement variables $\bm{\xi}_p$, the exact ground-state energy can be computed by diagonalizing the $3N \times 3N$ ($N$ being the number of atoms) matrix comprised of $3\times3$ matrices describing the coupling between each pair of atoms $i$ and $j$:
\begin{equation} \label{Cij}
\mathbf{C}^{\rm{MBD}}_{ij}=\delta_{ij}\omega_i^2+(1-\delta_{ij})\omega_i\omega_j\sqrt{\alpha_i \alpha_j}\mathbf{T}_{ij} \, .
\end{equation}
The diagonalization of the $\mathbf{C}^{\rm{MBD}}$ matrix leads to a set of $3N$ eigenvalues $\lambda_i$, corresponding to the squared frequencies of $3N$ ``dressed'' QHO modes. The interaction energy can then be computed as the difference between the interacting and non-interacting frequencies:
\begin{equation}
\label{EcQHO}
E_{\rm{c, MBD}}=\frac{1}{2}\sum_{i=1}^{3N} \sqrt{\lambda_i} - \frac{3}{2}\sum_{j=1}^N \omega_j \, .
\end{equation}

As already mentioned, under the condition that the dipole-dipole interaction $\mathbf{T}$ is frequency independent, the RPA correlation energy for the MBD Hamiltonian exactly equals the full MBD interaction energy.~\cite{Tkatchenko_2013a} As a direct consequence, the MBD model embodies a highly efficient tool for an effective RPA computation of the long-range correlation energy in the dipole limit. We also remark that the present diagonalization procedure represents a significant simplification with respect to the direct application of the ACFD formula, as it avoids the explicit integration over the frequency variable.

For solids there are two additional considerations required for accurately determining the MBD energy. First, the reciprocal space of the explicit dipole-dipole interactions must be sampled. This can be achieved by performing the MBD calculation on a supercell of the simulation cell. For molecular crystals these supercells are typically of $\sim$30~\AA ~in size in each direction, which is comparable to the $k$-point sampling of the underlying DFT calculations.~\cite{Reilly_2013a,Reilly_2013c} Second, to account for the interactions of the periodic neighbors of the atoms in the supercell we modify the expression for the elements of the $\mathbf{C}^{\rm{MBD}}$ matrix as follows:
\begin{equation} \label{Cij_solid}
\mathbf{C}^{\rm{MBD}}_{ij}=\delta_{ij}\omega_i^2+(1-\delta_{ij})\omega_i\omega_j\sqrt{\alpha_i\alpha_j}(\mathbf{T}_{ij}+\sum_{j'} \mathbf{T}_{ij'}) \, ,
\end{equation}
where the summation over $j'$ denotes that we consider coupling between $i$ and all of the periodic images of the $j$ within a certain spherical cut-off radius (which is typically $\sim$ 80~\AA). As noted in the main text, a larger cut-off radius was used for graphite due to the very long-ranged nature of the screening behavior in the constituent graphene sheets.~\cite{Gobre_2013}

\begin{acknowledgments}
Financial support from the European Research Council (ERC Starting Grant \texttt{VDW-CMAT}) is acknowledged.
\end{acknowledgments}

\bibliography{./literature_mbd}

\end{document}